# SuperCam, a 64-pixel heterodyne imaging array for the 870 micron atmospheric window

Christopher Groppi[*a], Christopher Walker[a], Craig Kulesa[a], Patrick Pütz[a,b], Dathon Golish[a], Paul Gensheimer[a], Abigail Hedden[a], Shane Bussmann[a], Sander Weinreb[c,d], Thomas Kuiper[c], Jacob Kooi[d], Glenn Jones[d], Joseph Bardin[d], Hamdi Mani[d], Arthur Lichtenberger[e], Gopal Narayanan[f]

[a]Steward Observatory, University of Arizona, 933 N. Cherry Ave., Tucson, AZ 85721 USA
[b]I. Physikalisches Institut Universität zu Köln Zülpicher Straße 77 50937 Köln, Germany
[c]NASA Jet Propulsion Laboratory, 4800 Oak Grove Dr., Pasadena, CA, 91109, USA
[d]California Institute of Technology, 1200 E. California Bvld., Pasadena, CA, 91125, USA
[e]University of Virginia 351 McCormack Rd., Charlottesville, VA, 22904, USA
[f]University of Massachusetts, 710 N. Pleasant St, Amherst, MA 01003, USA

## ABSTRACT

We report on the development of *SuperCam*, a 64 pixel, superheterodyne camera designed for operation in the astrophysically important 870 μm atmospheric window. *SuperCam* will be used to answer fundamental questions about the physics and chemistry of molecular clouds in the Galaxy and their direct relation to star and planet formation. The advent of such a system will provide an order of magnitude increase in mapping speed over what is now available and revolutionize how observational astronomy is performed in this important wavelength regime.

Unlike the situation with bolometric detectors, heterodyne receiver systems are coherent, retaining information about both the amplitude and phase of the incident photon stream. From this information a high resolution spectrum of the incident light can be obtained without multiplexing. *SuperCam* will be constructed by stacking eight, 1x8 rows of fixed tuned, SIS mixers. The IF output of each mixer will be connected to a low-noise, broadband MMIC amplifier integrated into the mixer block. The instantaneous IF bandwidth of each pixel will be ~2 GHz, with a center frequency of 5 GHz. A spectrum of the central 500 MHz of each IF band will be provided by the array spectrometer. Local oscillator power is provided by a frequency multiplier whose output is divided between the pixels by using a matrix of waveguide power dividers. The mixer array will be cooled to 4K by a closed-cycle refrigeration system. *SuperCam* will reside at the Cassegrain focus of the 10m Heinrich Hertz telescope (HHT). A prototype single row of the array will be tested on the HHT in 2006, with the first engineering run of the full array in late 2007. The array is designed and constructed so that it may be readily scaled to higher frequencies.

## 1. INTRODUCTION

SuperCam will operate in the astrophysically rich 870 micron atmospheric window, where the HHT has the highest aperture efficiency of any submillimeter telescope in the world and excellent atmospheric transmission more than 40% of the time. The proposed Superheterodyne Camera (SuperCam) will be an 8 x 8, integrated receiver array fabricated using leading-edge mixer, local oscillator, low-noise amplifier, cryogenic, and digital signal processing technologies.

SuperCam will be several times larger than any existing spectroscopic imaging array at submillimeter wavelengths. The exceptional mapping speed that will result, combined with the efficiency and angular resolution achievable with the HHT, will make SuperCam the most uniquely-powerful instrument for probing the history of star formation in our Galaxy and the distant Universe. SuperCam will be used to answer fundamental questions about the physics and chemistry of molecular clouds in the Galaxy and their direct relation to star and planet formation. Through Galactic surveys, particularly in CO and its isotopomers, the impact of Galactic environment on these phenomena will be realized. These studies will serve as "finder charts" for future focused research (e.g. with ALMA) and markedly improve the interpretation, and enhance the value of numerous contemporary surveys.

[*]cgroppi@as.arizona.edu; phone 1 520 626 1627; fax 1 520 621 1532

# 2. SUPERCAM SCIENCE

From the Milky Way to the highest-redshift protogalaxies at the onset of galaxy formation, the internal evolution of galaxies is defined by three principal ingredients that closely relate to their interstellar contents:

1. The transformation of neutral, molecular gas clouds into stars and star clusters (star formation).
2. the interaction of the interstellar medium (ISM) with the young stars that are born from it, a regulator of further star formation.
3. the return of enriched stellar material to the ISM by stellar death, eventually to form future generations of stars.

The evolution of (the stellar population of) galaxies is therefore determined to a large extent by the life cycles of interstellar clouds: their creation, starforming properties, and subsequent destruction by the nascent stars they spawn. The life cycle of interstellar clouds is summarized pictorially in Figure 1. Although these clouds are largely comprised of neutral hydrogen in both atomic and molecular form and atomic helium, these species are notoriously difficult to detect under typical interstellar conditions. Atomic hydrogen is detectable in cold clouds via the 21 cm spin-flip transition at 1420 MHz, but because the emission line is insensitive to gas density, cold (T~70K) atomic clouds are not distinguishable from the warm (T~8000K) neutral medium that pervades the Galaxy. Furthermore, neither atomic helium nor molecular hydrogen ($H_2$) have accessible emission line spectra in the prevailing physical conditions in cold interstellar clouds. Thus, it is generally necessary to probe the nature of the ISM via rarer trace elements. Carbon, for example, is found in ionized form ($C^+$) in neutral HI clouds, eventually becoming atomic (C), then molecular as carbon monoxide (CO) in dark molecular clouds. The dominant ionization state(s) of carbon accompany each stage of a cloud's life in Figure 1. In general, however, only global properties can be gleaned from the coarse spatial resolution offered by studies of external galaxies. Therefore detailed interstellar studies of the widely varying conditions in our own Milky Way Galaxy serve as a crucial diagnostic template or "Rosetta Stone" that can be used to translate the global properties of distant galaxies into reliable estimators of star formation rate and state of the ISM. These studies are very incomplete, however. Though we are now beginning to understand star formation, the formation, evolution and destruction of molecular clouds remains shrouded in uncertainty. The need to understand the evolution of interstellar clouds as they directly relate to star formation has become acute. The National Research Council's most recent Decadal Survey, under the advisory of a distinguished committee, has identified the study of star formation as one of the key recommendations for new initiatives in this decade. Similarly, understanding the processes that give rise to star and planet formation represent the central theme of NASA's ongoing Origins program. A new, comprehensive survey of the Galaxy must address the following questions to make significant progress toward a complete and comprehensive view of Galactic star formation:

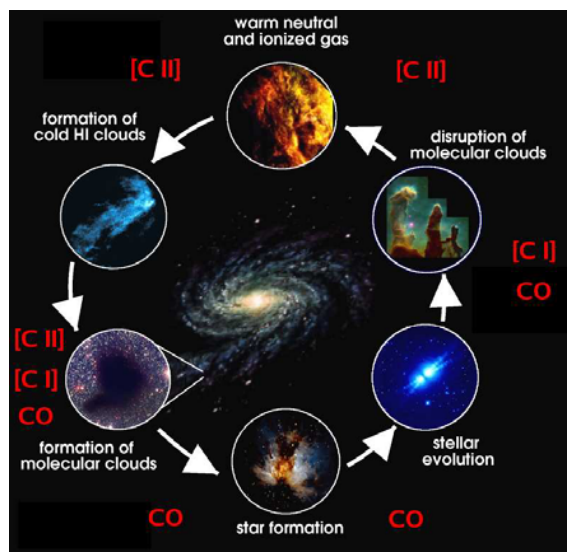

Figure 1: Life cycle of the ISM

- How do molecular clouds form, evolve, and get disrupted? How do typical atoms and grains cycle through the ISM?
- How and under what conditions do molecular clouds form stars?
- How do the energetic byproducts of stellar birth, UV radiation fields and (bipolar) mass outflows regulate further star formation in molecular clouds?
- How does the Galactic environment impact the formation of clouds and stars? What are the specific roles of spiral arms, central bars, and infall and other influences from outside the Galaxy?

## 2.1. SUPERCAM GALACTIC PLANE SURVEY

The following features represent a definitive survey that would not only provide the clearest view of the star forming clouds in the Galaxy, but would also serve as the reference map for future focused studies with the LMT, and SMA, CARMA and ALMA interferometers.

### 2.1.1. High Resolution Spectroscopic Imaging

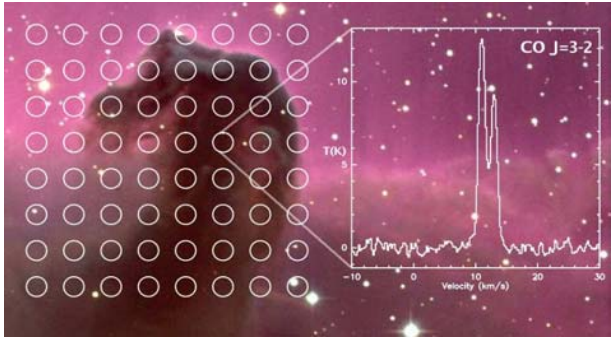

Figure 2: The 64 beams of SuperCam overlaid upon the Horsehead Nebula (B33). Each beam will measure a high resolution spectrum, a small portion (15%) of which is shown at right. The depicted spectrum was taken with the existing 345 GHz facility receiver at the HHT, and resampled to match the spectral resolution of SuperCam.

Techniques commonly used to diagnose the molecular ISM include submillimeter continuum mapping of dust emission (Hildebrand, 1983) and dust extinction mapping at optical and near-infrared wavelengths (Lada, Lada, Clemens, & Bally, 1994). Large format detector arrays in the infrared are now commonplace, and with the advent of bolometer arrays like SCUBA at the JCMT and SHARC at the CSO, both techniques have performed degree-scale maps of molecular material. However, these techniques have limited applicability to the study of the large-scale evolution of molecular clouds due to the complete lack of kinematic information. The confluence of many clouds along most Galactic lines of sight can only be disentangled with spectral line techniques. Fitting to a model of Galactic rotation is often the only way to determine each cloud's distance and location within the Galaxy. With resolution finer than 1 km/s, a cloud's kinematic location can be even distinguished from other phenomena that alter the lineshape, such as turbulence, rotation, and local effects such as protostellar outflows. These kinematic components play a vital role in the sculpting of interstellar clouds, and a survey that has the goal of understanding their evolution must be able to measure them. CO is second only to $H_2$ as the most abundant molecule in the ISM, and it remains the most accurate, most sensitive tracer of $H_2$ on large scales (Figure 2). The proposed SuperCam instrument will resolve the intrinsic profiles of Galactic CO lines, with a per-channel resolution of 0.2 km/s over 230 km/s of spectrometer bandwidth, comparable to the Galactic rotational velocity.

### 2.1.2. First Submillimeter Galactic Plane Survey

Molecular line surveys have been performed over the entire sky in the light of the 2.6 millimeter J = 1-0 line of $^{12}CO$, and have been used to synthesize our best understanding of the molecular content of the Galaxy. Still, our understanding of Galactic molecular clouds is incomplete. Early results were obtained with large beams, e.g., >9' (Dame et al., 1987; Dame, Hartmann, & Thaddeus, 2001)); were undersampled, e.g., 3' for the UMass/Stonybrook survey - (Solomon, Rivolo, Barrett, & Yahil, 1987; Scoville et al., 1987); or had limited areal coverage, e.g., the early FCRAO surveys - (Carpenter, Snell, & Schloerb, 1995; Stark & Brand, 1989; Bally, Langer, & Liu, 1991; Miesch & Bally, 1994). The Galactic Ring Survey (GRS) at FCRAO is by far the most comprehensive survey of the inner Galaxy to date (Simon et al., 2001). However, this survey traces only the J = 1-0 line of $^{13}$CO, which is less sensitive to warm, low-opacity, high velocity gas such as produced by outflows, photodissociation regions (PDRs), and shocks. This point is illustrated in Figure 3, with images of a synthetic model cloud constructed in the integrated light of different spectral lines of CO. The model cloud is externally illuminated by a B star and cloud excitation, temperature and chemical abundances are determined selfconsistently using Monte Carlo methods. The integrated spectral line images show that the heated portion of the cloud is largely missed by the J=1-0 lines, but captured by the J=3-2 lines. Reconstruction of the cloud based on observation of the $^{13}$CO J=1-0 line alone recovers only 60% of the total cloud mass, whereas the combination of J=1-0 and J=3-2 lines recovers 90% of the $H_2$ mass. A more comprehensive view of molecular clouds can therefore be gleaned from measurement of the submillimeter lines of CO and its isotopes, in combination with existing millimeter-wave observations. The gas probed by higher-J transitions is of greatest interest to our posed questions - it is the energetic gas that 1) participates in molecular outflows, 2) senses radiation fields at the photodissociated surfaces of clouds, and 3) is warmed by starformation in cloud cores. Higher-J lines are also needed to properly interpret even basic properties of clouds derived from existing CO 1-0 observations. Due to the prevailing physical conditions in the interstellar medium, the 870 micron (320-370 GHz) atmospheric window is one of the richest in the electromagnetic spectrum (Figure 4-a). This window also has the highest atmospheric transmission of any submillimeter band. At this wavelength the HHT has the highest aperture effciency (80%) of any submillimeter telescope in the world and excellent atmospheric transmission more than 40% of the time (Figure 4-b). These prospects make the design of a large format multibeam receiver in the 345 GHz atmospheric window most attractive.

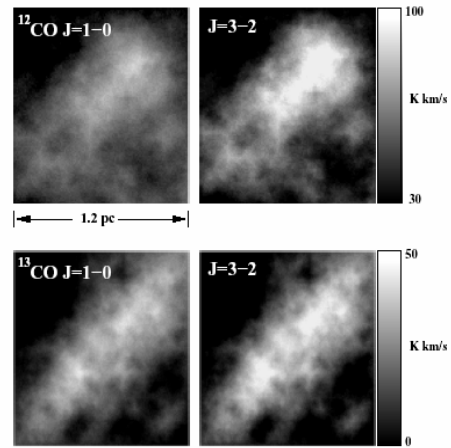

Figure 3: Simulated image of a fractal molecular cloud in several CO transitions. The energetic gas that interacts with stars is far better probed by the 3-2 line, however both 1-0 and 3-2 lines are needed together to extract a comprehensive understanding of cloud properties, dynamics & evolution.

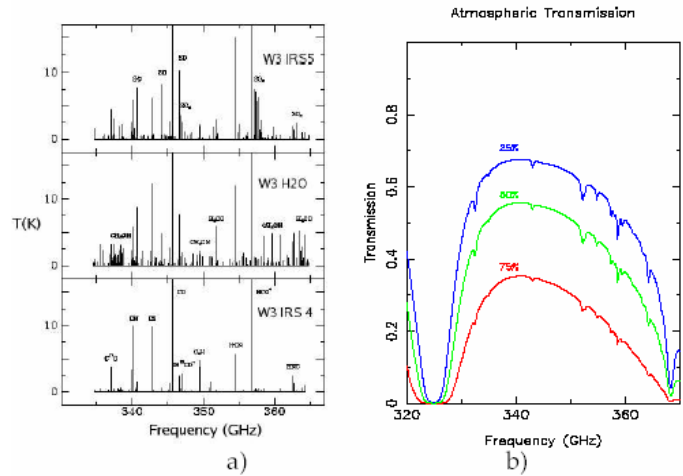

Figure 4: a) The 345 GHz spectral line survey toward three positions in the W3 molecular cloud by Helmich & van Dishoeck (1997) shows a rich diversity of spectral diagnostics. b) Modeled submillimeter atmospheric transparency for the HHT on Mt. Graham in 75 percentile (top), median (middle), and 25 percentile (bottom) atmospheric conditions, derived from 24 hour 225 GHz radiometer measurements over the last 6 years.

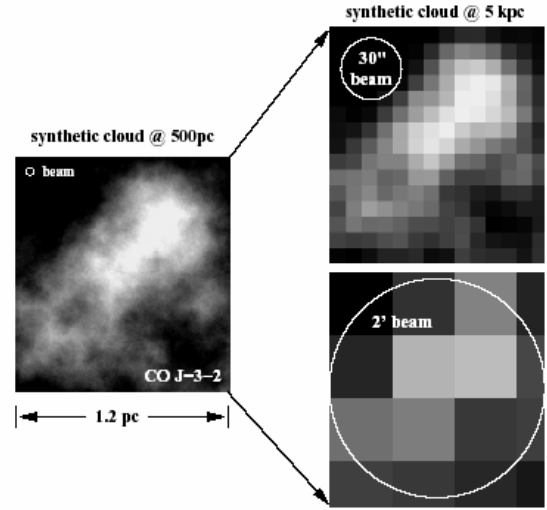

Figure 5: : The synthetic cloud from Figures 5 as seen in the 12CO J=3-2 transition at a distance of 500 pc (left) and at 5 kpc, with beam sizes of 30" (top) and 2' (bottom). The structure of the cloud is essentially lost in the larger beam. In order to probe cloud structure and excitation over the entire Galactic disk, high angular resolution is vital.

### 2.1.3. High Angular Resolution

Angular resolution is a critical aspect of improvement for a new Galactic survey. Figure 5 depicts the model cloud of Figure 3, projected to distances of 500 pc and 5 kpc. Clearly, disentangling different clouds and cloud components can only be accomplished with sub-arcminute angular scales. The angular resolution of SuperCam at the HHT is 23" at 345 GHz.

### 2.1.4. High Sensitivity

CO survives in the ISM in part because of the UV shielding from dissociation provided by $H_2$; thus CO's survivability depends upon a molecular, $H_2$-dominated environment. For typical molecular clouds, the sharp transition from H to $H_2$ typically occurs by a visual extinction of ~1 magnitude in the local interstellar radiation field, or $N(H) = 1.8 \times 10^{21}$ $cm^2$. We therefore aim to detect all CO down to this hydrogen column density limit. This corresponds to a $3\sigma$ detection limit of $N(^{12}CO) \sim 10^{15}$ $cm^2$, which implies an integrated intensity for cold gas ($10K < T_k < 50K$) of 1.2 K km/s in the J = 3-2 transition at a gas density of $n_H = 10^4$ $cm^3$. This sensitivity limit is achievable ($3\sigma$) within 10 seconds of integration time per independent beam in *median* atmospheric conditions (Tsys ~700K) at the HHT, or (100/#pixels) hours per mapped square degree, with 10" grid spacing. Detection (or limits) on J=3-2 in that time would constrain the gas density, based upon the line brightness of millimeter wave transitions.

### 2.1.5. Mapping Coverage of the Galactic Plane

Figure 6 demonstrates the needed sky coverage of a submillimeter-wave Galactic plane survey. From previous CO surveys it is known that the scale height of CO emission toward the inner Galaxy is less than one degree (Dame et al., 1987; Dame, Hartmann, & Thaddeus, 2001). The interstellar pressure, abundances, and physical conditions vary strongly as a function of Galactocentric radius, so it is necessary to probe the inner Galaxy, the outer Galaxy, and the l±100° tangent arms to obtain a statistically meaningful survey that encompasses the broad dynamic range of physical conditions in the Galaxy. We will therefore probe the entire Galactic plane as seen from Arizona (0 < l < 240°). Below l = 90°, a *completely unbiased survey* will be undertaken, covering 180 square degrees (1° < b < 1°). This "inner" Galaxy survey will coincide with two synergistic surveys: the FCRAO-BU Galactic Ring Survey (GRS) and GLIMPSE, a Spitzer Space Telescope (SST) Legacy Program (Benjamin et al., 2003). Above l = 90°, most of the CO emission is located at higher Galactic latitude, so the same solid angle will be distributed according to the Dame et al. (1987); Dame, Hartmann, & Thaddeus (2001) survey to follow the CO 1-0 distribution and the best characterized star forming regions in the Galaxy, while maximizing synergies with the "Cores to Disks" SST Legacy program, and other SST GTO programs.

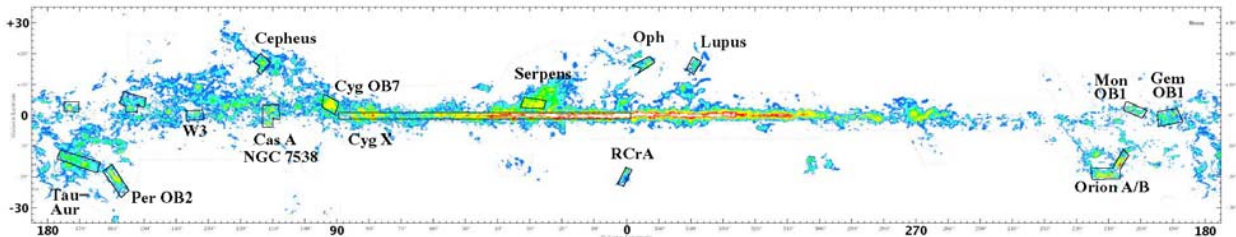

Figure 6: The power of SuperCam: a definitive chemical and kinematic survey of star forming clouds in $^{12}CO$ & $^{13}CO$ J=3-2 over 500 square degrees ($^{12}CO$) of the sky can be performed in 33 full days per spectral line. A corresponding survey with contemporary single pixel receivers would take over 6 years each.

### 3. SUPERCAM INSTRUMENT DESCRIPTION

In order to achieve our scientific objectives we are building the first large format (64 pixel) superheterodyne camera (SuperCam) at submillimeter wavelengths. The advent of such a system will provide an order of magnitude increase in mapping speed over what is now available and revolutionize how observational astronomy is performed in this important wavelength regime. Unlike the situation with bolometric detectors, heterodyne receiver systems are coherent, retaining information about both the amplitude and phase of the incident photon stream. From this information a high resolution spectrum of the incident light can be obtained without multiplexing. Indeed, **each** SuperCam pixel will provide 2,048 simultaneous spectral measurements. In terms of raw power, each observation made with *SuperCam* will provide 131,072 independent measurements of the properties of the object under study. High resolution spectroscopy can, in principle, be performed in this same wavelength regime using incoherent detectors together with frequency dispersive quasioptical devices such as gratings and Fabry-Perot interferometers. However, the size requirement of quasi-optical devices and/or the need to scan in order to construct a spectrum make them too cumbersome or insensitive for the scientific objectives of the proposed study. The possibility of large format arrays at submillimeter wavelengths has been discussed for more than two decades (Gillespie & Phillips, 1979), but was prohibited by several factors:

- Sensitive mixing devices either did not exist or were difficult to fabricate.
- When mixers were available, their performance would often vary significantly from device to device.
- There was insufficient LO power to simultaneously pump more than one or two detectors.
- Stacking more than a few mixer blocks together in the focal plane with their associated backshorts (if necessary), IF amps, magnets, bias lines, etc. was mechanically complex and could overload the cryogenic system.
- The cost of the frontend components and the required backend spectrometer were prohibitive.

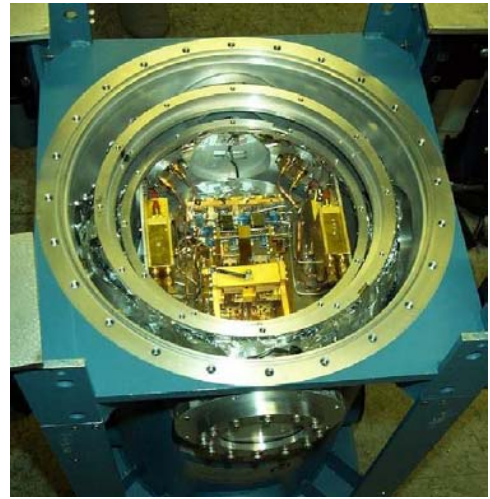

Figure 7: Inside the PoleStar dewar.

However, through the arduous efforts of many researchers in the field, these hurdles have now been overcome. The technology developed in this effort has applications in a number of research areas outside of astronomy. These include remote imaging, space-based communications, and hazardous material detection.

### 3.1. INSTRUMENT DESIGN

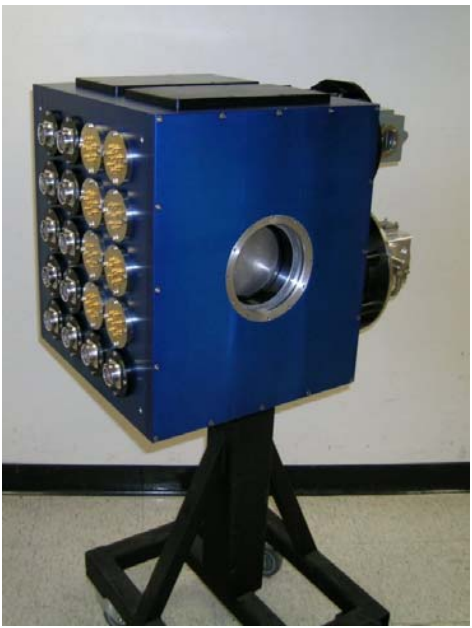

Figure 8: The SuperCam cryostat

Unlike all other millimeter/submillimeter arrays composed of individual mixers and discrete components, the array has a high degree of integration. Well conceived, efficient packaging is essential to the successful implementation of large format systems. Figure 7 shows the crowded interior of the 4 pixel (*PoleSTAR*) array cryostat built by the University of Arizona from discrete components. The enormous complexity of even a small discrete system suggests a more integrated approach for larger systems. At the heart of the array is an 8x8 integrated array of low-noise mixers. The array mixer contains first stage, low-noise, MMIC IF amplifier modules with integrated bias tees. A single solid-state source provides local oscillator power to each array mixer via a waveguide corporate power divider and a simple Mylar diplexer. SuperCam is fully automated, with a field-proven control system that provides the user with a full range of instrument control and characterization options. Below we discuss SuperCam's key components.

#### 3.1.1. CRYOGENICS

The SuperCam cryostat is shown in Figure 8. The cryostat's outer dimensions are 546x546x481 mm. Light from the telescope enters the cryostat through a 127 mm (500) diameter AR coated, crystalline quartz vacuum window and passes through an IR blocking filter on the 40 K radiation shield before illuminating the 4 K mixer array. SuperCam uses a Sumitomo Model a SRDK-415E-A71A cryocooler. The cooler has 1.5 W of thermal load capacity at 4.2 K and 45W at 40K with orientation-independent operation. The operating temperature of the cryocooler is stabilized by the addition of a helium gas pot on the 2nd stage. Once the 2nd stage cools to 4 K, the helium gas liquifies. The SuperCam coldplate is heatsunk to this pot via low-loss, vibration-damping copper straps. A CTI cryogenics CTI-350 coldhead supplements the cooling of the 40K shield, and provides 12K heatsinking for the 64 stainless steel semi-rigid IF cables. The addition of this second coldhead permits the use of moderate lengths of standard coaxial cable while maintaining low heat load at 4K. Calculations indicate the SRDK-415E load capacity is sufficient to cool the mixers, magnets, and amplifiers to the proper operating temperatures. The cryostat was constructed by Universal Cryogenics in Tucson, Arizona, USA. Cryogenic load testing using heaters at all temperature stages have verified that both the cryocoolers meet their rated load specifications. The cooling capacity is adequate given the expected heat loading from the DC wiring, semi-rigid cable, amplifiers and magnets, with an expected load capacity margin of ~50%.

### 3.1.2. MIXER ARRAY

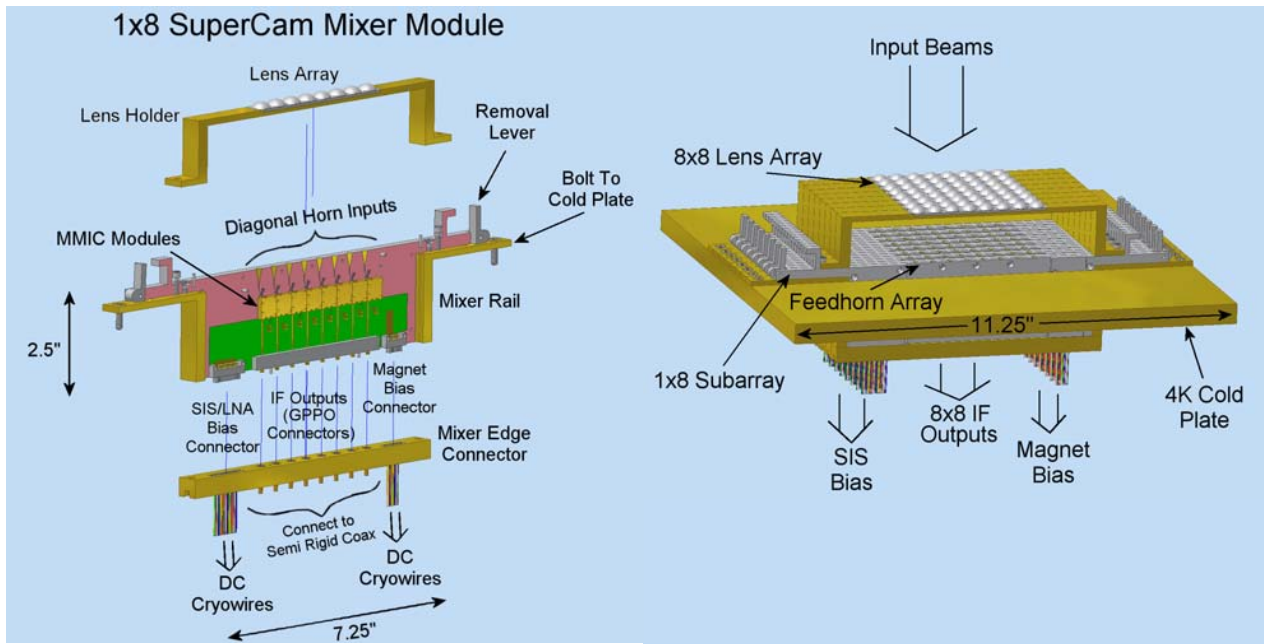

Figure 9: Assembly diagram of a single mixer module.          Figure 10: Assembly diagram of the complete array

In the 300-400 GHz range no amplifiers exist, so incoming signals must first be down-converted to frequencies where low noise amplifiers are available. At these high frequencies two types of mixing devices are currently in use; Schottky diodes and superconductor-insulator-superconductor (SIS) junctions. Schottky diode mixers have been in use on ground, airborne, and space-based platforms for many years. They are reliable and provide modest sensitivity, which improves as they are cooled. However, they require prodigious amounts (of order mW) of local oscillator (LO) power to efficiently pump a single mixer. State-of-the-art SIS mixers in the 350 GHz band have 20 times the sensitivity of Schottky mixers. Their greater sensitivity and relatively low LO power requirement make SIS mixers the best choice for array applications. To operate, these mixers must be cooled to ~4 K. We are developing a compact, sensitive, 64 pixel array of SIS mixers optimized for operation in the 320-360 GHz atmospheric window. The two dimensional, 8x8 array will be composed of eight, 1x8 subarrays. The array mixers will utilize SIS devices fabricated on Silicon-On-Insulator (SOI) substrates, with beam lead supports and electrical contacts. The waveguide probe and SIS junction are based on an asymmetric probe design currently in use at the Caltech Submillimeter Observatory in their new facility 350 GHz receiver. The measured DSB noise temperature of this receiver (40 K) is excellent and essentially frequency independent across the band. The 1x8 mixer subarrays will be constructed from tellurium copper using the splitblock technique. Stainless steel guide pins and screws are used to ensure proper alignment and good contact between parts. An assembly diagram is shown in Figure 9. Figure 10 is a pictoral representation of one mixer in a 1x8 subarray. A low-loss, dielectric lens couples energy from the telescope into a diagonal feedhorn. This type of horn offers excellent Gaussian beam coupling at the expense of significant cross-polarization. Since SuperCam is a single polarization design with no polarization selective optics, the cross-polarization introduced by the horns will not impact the instrument performance. The energy in the horn passes through a 90° waveguide bend before reaching the impedance transformer to full height rectangular waveguide. The SIS device is suspended above the suspended stripline channel via four small beamlead supports. The device is secured in place with Crystalbond adhesive. Both the hot and ground beamleads are tack-bonded with a wirebonder to the MMIC module input pad and block, respectively. The mixer blocks will be fabricated at the University of Arizona using a Kern MMP micromilling machine purchased for this project. This numerically controlled mill can fabricate structures to micron accuracy with a high level of automation. A 24 position automated tool changer, touch probe system and laser tool metrology system allow the fabrication of a complete mixer block from raw stock in a small number of steps with a minimum of user intervention.

### 3.1.3. Local Oscillator

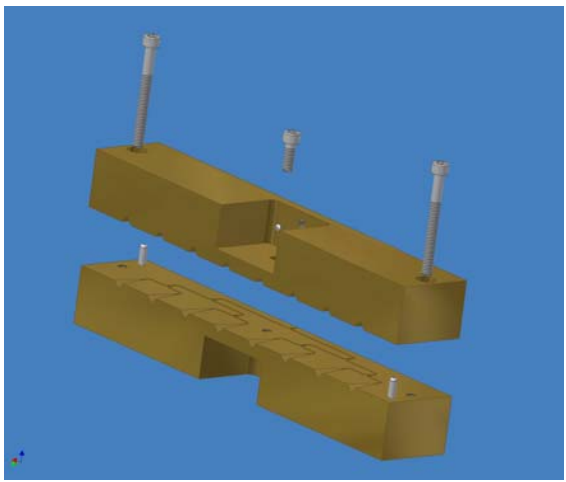
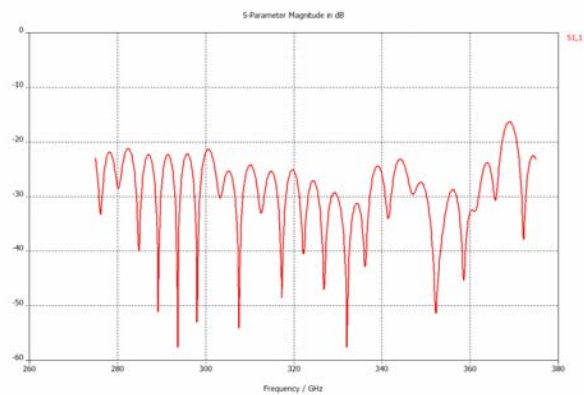

Figure 11: 8-way corporate power divider and simulated performance including losses from finite conductivity and surface roughness.

With an array receiver, LO power must be efficiently distributed among pixels. Depending on the mechanical and optical constraints of the array, a balanced distribution can be achieved using quasioptical techniques or waveguide injection. With the quasioptical approach, dielectric beam splitters or holographic phase gratings are used to divide the LO energy between array pixels. The multiple LO beams are then combined with the corresponding sky beams before entering the array mixers using a beam splitter, Martin-Puplett interferometer, or Fabry-Perot interferometer. The quasioptical approach works well for modest sized arrays. However, for the large format system being proposed here, the size of the required quasi-optical power splitter and diplexer become prohibitive. Therefore we have chosen to use a hybrid waveguide/quasioptical LO power injection scheme. The LO power for the array will be provided by a single solid-state, synthesizer-driven source available from Virginia Diode Inc. The active multiplier chain consists of a high power solid-state amplifier followed by a series of tunerless broadband multipliers. There are no mechanical tuners, so the output frequency simply tracks the synthesized input frequency. The chain utilizes a series of broadband varactor doublers that have been developed at Virginia Diode Inc. These doublers have a tunerless bandwidth of 15-20% and exhibit efficiencies from 60% at 50 GHz to 10-20% at 300 GHz. The output of the array is coupled to an eight-way waveguide corporate power divider with splitblock machineable waveguide twists. Each of the eight outputs provides the drive power for a 1x8 subarray via an identical 8 way corporate divider with diagonal waveguide feedhorn outputs. Figure 11 shows a 3D model of a 8 way power divider, and performance simulated with CST Microwave Studio. A lens array similar to the mixer lens system then matches the LO beams to the mixers through a simple Mylar beamsplitter diplexer. This scheme ensures uniform LO power in each beam since the waveguide path lengths are identical for each beam. In addition, the waveguide feedhorns and lenses provide well controlled and predictable LO power distribution and coupling to each mixer. Accounting for conduction and surface roughness losses, we expect this 64-way network to add an additional 2dB of LO power loss compared to a lossless divider.

### 3.1.4. IF/BIAS DISTRIBUTION SYSTEM

#### 3.1.4.1. IF AMPLIFIERS

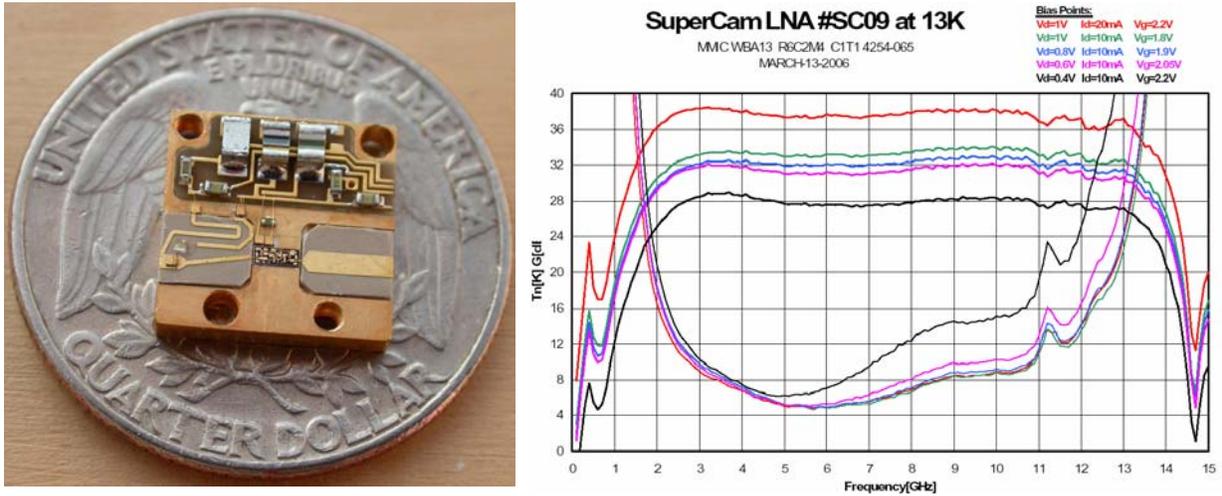

Figure 12: A SuperCam MMIC amplifier module, and typical measured results at 13K bath temperature for several bias points. Amplifier noise remains low for bias powers as low as 8 mW. Gain remains above 30 dB.

The IF outputs from the SIS devices are bonded directly to the input matching networks of low-noise, InP MMIC amplifier modules located in the array mixers. These amplifier modules have been designed and fabricated by Sander Weinreb's group at Caltech. The IF center frequency of the array will be 5 GHz. This is the standard IF frequency at the HHT and lies within the operating range of the existing generation of IF amplifier modules. During the past 4 years wideband, very low noise, cryogenic monolithic microwave integrated circuit (MMIC) amplifiers have been developed with design and testing at JPL and Caltech and foundry fabrication at TRW (now Northrop Grumman Space Systems, NGST) and HRL. These LNA's utilize indium phosphide, 0.1 micron gate length, high electron-mobility transistors (HEMT's) and match the requirements needed for densely packed focal plane arrays in terms of noise temperature, chip size, DC power dissipation, yield, and bandwidth. The 1 to 12 GHz chip is most appropriate for this application because of its match to the IF bandwidth of SIS and HEB devices and its very low noise and power consumption. The chip is contained in an 11mm x 11mm amplifier module that contains integrated bias tees for the SIS device and the amplifier chip. The module achieves noise temperature of ~5 K consuming 8 mW of power at 4K. The first 10 amplifier modules are complete. An example is shown in figure 12, with measured gain and noise data at 8 mW power dissipation. We have integrated an amplifier module into a single pixel SIS mixer and have verified that the amplifier module operates as expected. Allan varience times and mixer noise temperatures are unchanged within the measurement errors compared to a similar mixer used with an external commercial LNA and cryogenic isolator.

#### 3.1.4.2. ARRAY SPECTROMETER

The original baseline design for the SuperCam spectrometer was a 64-channel autocorrelator with 256 MHz analog bandwidth and 0.3 MHz resolution (220 km/s and 0.25 km/s). Today, the SuperCam spectrometer will deliver 64 channels at 512 MHz/channel with 125 kHz resolution, or 32 channels at 1 GHz with 250 kHz resolution, all for the same price. This technological advancement allows SuperCam to more thoroughly execute the Galactic plane survey project. The system will be capable of resolving lines in the coldest clouds, while fully encompassing the Galactic rotation curve. Operated in 32 pixel mode with 1 GHz (880 km/s) of bandwidth, extragalactic studies are enabled. This leap in spectrometer ability is driven by the rapid expansion in the capabilities of high speed Analog to Digital Converters (ADCs) and Field Programmable Gate Arrays (FPGAs). The SuperCam spectrometer, to be provided by Omnisys AB of Sweden, is based on a real-time FFT architecture. High speed ADCs digitize the incoming RF signal at greater than 10 bits resolution, preventing any significant data loss as with autocorrelation based schemes. Then, large, high speed FPGAs perform a real time FFT on the digitized signal and integrate the incoming signal. Only recently has Xilinx released FPGAs fast enough and large enough to accommodate the firmware capable of this task. These systems are fully reconfigurable by loading new firmware into the FPGAs. In addition, the spectrometer can be easily expanded to increase bandwidth. We will receive 16 boards capable of processing 64x512 MHz or 32x1 GHz. By adding 16 identical boards, we can increase the bandwidth processed to 64x1 GHz. Test results from the first SuperCam prototype spectrometer board are shown in figure 13.

### 3.1.5. OPTICS

The existing secondary mirror of the Heinrich Hertz Telescope provides a f/13.8 beam at the Nasmyth focus. The clear aperture available through the elevation bearing prevents the possibility of a large format array at this position. To efficiently illuminate a large format array like SuperCam, the telescope focus must fall within the apex room located just behind the primary. A system of re-imaging optics transforms the f number of the telescope to f/5. Since the physical separation between array elements in the instrument focal plane scales as $2f\lambda$, lower f/#'s serve to reduce the overall size of the instrument. The reimaging optics are composed of a hyperbola and an ellipse with two flat mirrors. All the reimaging optics can be mounted on a single optical breadboard and left in the apex room. The cryostat and optics frame have been designed using finite element analysis to minimize gravitational deflection, and the calculated deflections have been fed into the tolerancing of the optical design. The optical system was initially designed and optimized with Zemax, and was then verified by BRO research using their ASAP physical optics package. The system's efficiency exceeds 80% for all pixels, and has been verified to be robust to alignment and fabrication tolerances. The hyperbolic and elliptical mirror will be fabricated at the University of Arizona through direct machining with a CNC mill. A coordinate measuring machine (CMM) at the Steward Observatory Mirror Laboratory will be used to verify the figure of the completed optics. The optical system is depicted in Figure 14.

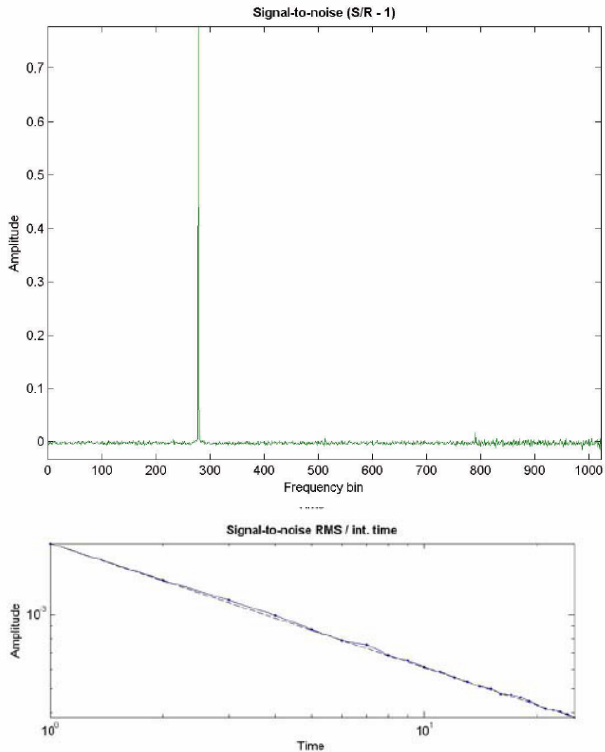

Figure 13: Laboratory spectrum and Allan variance plot from the first SuperCam prototype spectrometer board.

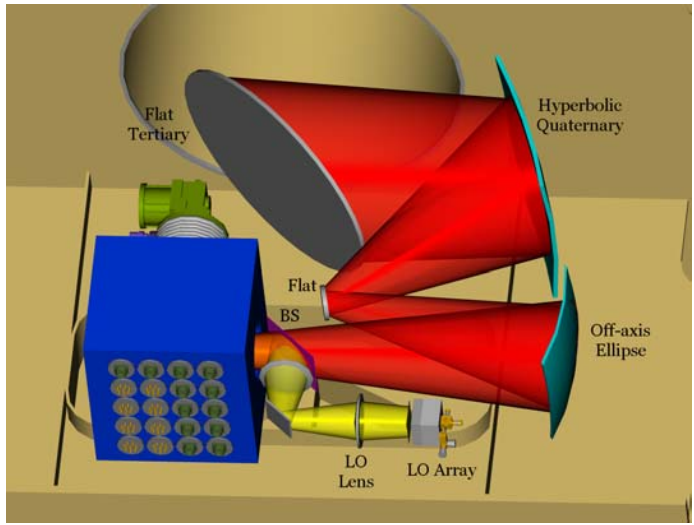

Figure 14: The SuperCam optical system.

## 4. CONCLUSION

We are constructing SuperCam, a 64-pixel heterodyne imaging spectrometer for the 870 micron atmospheric window. A key project for this instrument is a fully sampled Galactic plane survey covering over 500 square degrees of the Galactic plane and molecular cloud complexes. This $^{12}$CO(3-2) and $^{13}$CO(3-2) survey has the spatial (23") and spectral (0.25 km/s) resolution to disentangle the complex spatial and velocity structure of the Galaxy along each line of sight. SuperCam was designed to complete this survey in two observing seasons at the Heinrich Hertz Telescope, a project that would take a typical single pixel receiver system 6 years of continuous observing to complete. Prototypes of all major components have been completed or are being fabricated now. The first 1x8 mixer row will be integrated and tested in 2006, and the full array will be populated and tested in 2007. We expect first light observations in winter 2007.